\documentclass[]{spie}

\usepackage{amsmath,amsfonts,amssymb}
\usepackage{graphicx}
\usepackage[colorlinks=true, allcolors=blue]{hyperref}
\usepackage{subcaption}

\title{Inverse-designed tapers to improve throughput of on-chip astrophotonic spectrographs}

\author[a]{Vikram Seenivasan}
\author[a]{Pradip Gatkine}
\affil[a]{Department of Physics and Astronomy, University of California, Los Angeles, CA 90095}

\authorinfo{Further author information: 
\\Send correspondence to Vikram Seenivasan; e-mail ID: vikrams25@g.ucla.edu}

\begin{document} 
\maketitle

\begin{abstract}

Astrophotonics offers a pathway towards compact and stable spectrographs for future large telescopes. The size of these spectrographs does not scale with the diameter of the telescope aperture; thus, they offer a cost-effective way to achieve astronomy science requirements compared to traditional optical spectrographs. A major limitation of current on-chip astrophotonic spectrographs is the significant loss incurred at interfaces between distinct components, particularly at the input and output fiber interfaces, limiting overall throughput. In this work, we investigate inverse-designed architectures to optimize transitions between these components to improve throughput. Inverse design formulates device design as an optimization problem, enabling non-intuitive geometries that can achieve high performance. We focus on optimizing coupling between an arrayed waveguide grating (AWG) spectrograph and an input single-mode fiber. Our approach explores two classes of fabrication-constrained inverse design. The first is shape optimization, which optimizes the geometry of the boundary of a coupling taper. The second is topology optimization, which optimizes the material distribution in a design region to gradually transform the optical mode between components. Both methods are implemented using the Tidy3D package from Flexcompute. We use germanium-doped silica for the core of the photonic components and silica as the cladding, and we use an SMF-28 fiber as input. This work demonstrates a pathway to improve astrophotonic spectrographs through computational inverse-design methods. Leveraging higher throughput, we can improve signal-to-noise ratios offered by photonic devices and position ourselves to design high-performance, cost-effective instrumentation for the next generation of telescopes, both on the ground and in space.
\end{abstract}

\keywords{Inverse Design, Astrophotonics, Tapers, Astrophotonic Spectrographs, Arrayed Waveguide Gratings}

\section{INTRODUCTION}
\label{sec:intro}  

Upcoming ground-based astronomical telescopes are being built with increasingly large apertures to enable higher-precision observations for science cases ranging from cosmology to exoplanet detection \cite{TMT, ELTScienceCase, GMTScienceBook2018}. Several of these science goals rely on high-resolution spectroscopy, such as precision radial-velocity measurements, exoplanet atmospheric characterization, stellar chemical-abundance measurements, and tests of fundamental physics. However, scaling spectrographs to these next-generation telescopes presents a significant instrumentation challenge. Traditional bulk-optics spectrographs scale in both size and cost with telescope aperture ($D$)
as approximately $D^2$--$D^3$\cite{gatkine2019astrophotonicspectrographs}. Additionally, their increasing physical footprint makes the stringent thermal and mechanical stability requirements needed for high-resolution spectroscopy increasingly difficult to achieve \cite{chamarthi2018stability}.

The emerging field of astrophotonics offers a way to break the conventional scaling relation between telescope aperture and spectrograph size by enabling diffraction-limited operation\cite{2023AstrophotonicsRoadmap}. It does so by directly manipulating light on chips using technologies adapted from the telecommunications industry, such as optical fibers and photonic integrated circuits. This offers a pathway toward compact, stable, replicable, and potentially cost-effective spectrographs for future large telescopes. Their compactness and inherent stability also make them attractive for space-based telescopes, where instrument mass and volume are tightly constrained.

\begin{figure}
    \centering
    \includegraphics[width=0.8\linewidth]{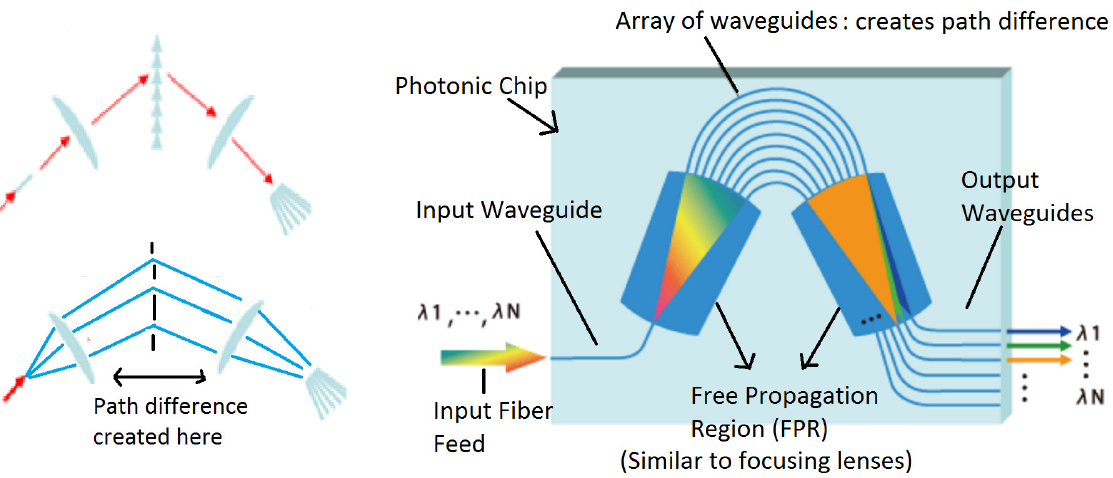}
    \caption{Comparison between a conventional optical grating spectrograph (left) and arrayed waveguide grating (right), reprinted with permission from [\citeonline{gatkine2019astrophotonicspectrographs}].}
    \label{fig:comparsionAWG}
\end{figure}

Arrayed waveguide gratings (AWGs) are a promising architecture for astrophotonic spectrographs\cite{gatkine2019astrophotonicspectrographs} and have been demonstrated on-sky for astronomical observations \cite{calderon_2026, Jovanovic_2017_demonstration, Pradeep_2025_demonstration, skorpen_2026}. Figure \ref{fig:comparsionAWG} compares a conventional optical grating spectrograph with an AWG architecture. 
A key challenge of current AWG-based astrophotonic spectrographs is achieving high overall throughput. This is crucial to achieve the high detector signal-to-noise ratios required for high-resolution astronomy science cases. A significant portion of light loss arises at interfaces between the different photonic components of the AWG, especially at the fiber-chip interface\cite{Gatkine_2016, 
2023AstrophotonicsRoadmap, gatkine2024efficientultrabroadbandlowresolutionastrophotonic}, where variations in modal profiles, device geometry, and material refractive indices lead to imperfect coupling. As AWGs contains multiple such interfaces, losses at these individual sites compound and can substantially reduce the intensity of transmitted light. Consequently, the development of high-throughput, broadband photonic couplers to interface between the AWG components is one way to progress these devices to practical application. 

Conventional forward-design approaches are limited by physical intuition and only explore a limited subset of the possible device design space. In this work, we investigate inverse design to optimize fiber-to-chip coupling of an AWG. Inverse design formulates device design as an optimization problem in which geometric or material parameters are iteratively updated using gradient-based methods to maximize a specified objective. Here, the objective is broadband optical throughput. Inverse design has been increasingly adopted across photonics, with applications including metasurfaces, fiber couplers, wavelength-division multiplexers, mode converters, nonlinear and quantum photonic devices, and other applications in silicon photonics \cite{schul2026inverse}. 

We present a preliminary study of inverse-designed  fiber-to-chip coupling using both shape and topology optimization, and briefly discuss the potential of inverse design for improving broadband throughput of future astrophotonic instrumentation.

\section{Methodology}

This section outlines the inverse-design methodology used in this study. We first describe the simulation setup followed by the inverse-design framework, including the objective used to evaluate device performance. We then describe the two parameterizations considered, shape optimization and topology optimization, and their corresponding optimization .

\subsection{Simulation setup}

In this work, we optimize the fiber-to-chip transition for a Ge-doped SiO$_2$ AWG similar to that from \citeonline{gatkine2024efficientultrabroadbandlowresolutionastrophotonic}. We adopt this material platform because it achieved the highest overall throughput among the devices investigated in that study. All electromagnetic simulations and gradient calculations are performed using the Tidy3D finite-difference time-domain (FDTD) and inverse-design framework \cite{tidy3d_github}.

\begin{figure}
    \centering
    \includegraphics[width=0.8\linewidth]{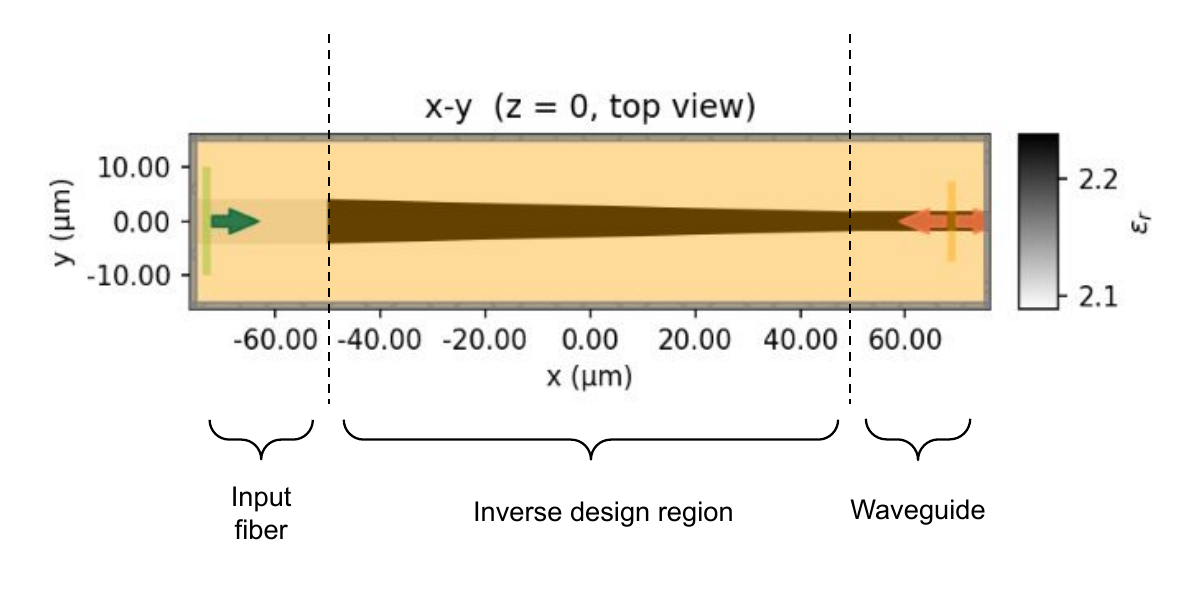}
    \caption{Simulation domain for the fiber-to-chip transition, shown in the propagation--transverse plane for shape optimization. From left to right, the domain consists of a $25~\mu\mathrm{m}$ section of SMF-28e+ fiber, a $100~\mu\mathrm{m}$ inverse-design region, and a $25~\mu\mathrm{m}$ section of uniform output waveguide. The ModeSource (green) excites the forward-propagating fundamental fiber mode, while the ModeMonitor (orange) measures the forward-propagating fundamental waveguide mode at the output. For topology optimization, the inverse-design region is instead represented by a rectangular pixel grid.}
    \label{fig:sim_figure}
\end{figure}

The simulation domain consists of the input fiber, the on-chip inverse-design region, and the on-chip waveguide, as shown in Figure~\ref{fig:sim_figure}. The input fiber is modeled using the specifications of Corning SMF-28e+ Photonic Optical Fiber \cite{CorningSMF28ePlus}, with an $8.2~\mu\mathrm{m}$ core diameter surrounded by silica cladding. A $25~\mu\mathrm{m}$ section of fiber precedes the inverse-design region. The inverse-design region is $100~\mu\mathrm{m}$ long and is followed by a $25~\mu\mathrm{m}$ section of uniform waveguide, which has a $3.6~\mu\mathrm{m}\times3.6~\mu\mathrm{m}$ cross section. The inverse-design region and output waveguide are surrounded by $15~\mu\mathrm{m}$ of silica cladding above and below, giving a total simulation domain of $150~\mu\mathrm{m}\times30~\mu\mathrm{m}\times33.6~\mu\mathrm{m}$ along the propagation, transverse, and vertical directions, respectively. Perfectly matched layer (PML) boundary conditions are applied on all sides of the simulation domain to absorb outgoing electromagnetic waves and prevent artificial reflections from contaminating the transmission calculation.

Due to computational limitations, material dispersion is neglected in this preliminary investigation. The refractive indices vary only modestly across the optimization range; for example, the refractive index of SiO$_2$ changes by approximately 0.4\% between 1200 and 1700~nm. The refractive indices are therefore held constant at values corresponding to the central wavelength of 1450~nm. The silica cladding is assigned a refractive index of 1.4452 from the Malitson dispersion relation \cite{Malitson:65}, while the SMF-28e+ core is assigned a refractive index of 1.4504 by linearly interpolating the manufacturer's specifications. The Ge-doped silica waveguide is assigned a refractive index of 1.4962, corresponding to an index contrast of approximately 3.5\% relative to the silica cladding.

\subsection{Inverse-design framework}

During the inverse-design process, gradient-based optimization is used to iteratively update a set of device parameters to maximize an objective function. This objective function depends on the device design parameters and quantifies the performance of a given design. The choice of device parameterization determines the range and type of device geometries that can be explored, as well as the degrees of freedom available to the optimizer. We investigate two complementary parameterizations for the inverse-design region: shape optimization and topology optimization. In shape optimization, the boundary of a taper connecting the fiber and waveguide is modified while its material composition remains fixed. In topology optimization, the material distribution throughout a prescribed design region is optimized, providing substantially greater geometric freedom. Shape and topology optimization are described in detail in Sections~\ref{subsec:shape_optimization} and~\ref{subsec:topology_optimization}, respectively.

The objective function consists of a performance term and a fabrication penalty. The performance term in the objective is the broadband fiber-to-waveguide coupling efficiency, defined as the fraction of input optical power coupled into the fundamental waveguide mode, evaluated over the 1200--1700~nm wavelength range. This wavelength range partially covers the astronomical J- (1170--1330~nm) and H-bands (1490--1780~nm), which are useful for applications including exoplanet characterization, and lies within one of the most technologically mature wavelength regimes for integrated photonics owing to developments in the telecommunications industry \cite{siliconphotonics}. Thus, it provides a useful wavelength range over which to investigate the performance limits of the fiber-to-chip transition. The fabrication penalties penalizes designs containing features smaller than the assumed minimum fabrication scale, steering the optimization toward more readily fabricable designs. These penalties are described separately for shape optimization in Subsection~\ref{subsec:shape_optimization} and topology optimization in Subsection~\ref{subsec:topology_optimization}. 

The optimization process consists of a series of iterative steps toward maximizing the objective function. At each step, the device is simulated and the objective function is evaluated. The device parameters are then updated using gradient-based optimization, and the process is repeated. In each simulation step, the fundamental mode of the SMF-28e+ fiber is excited using a broadband ModeSource centered at 1450~nm, with a Gaussian frequency spectrum spanning the entire optimization range. Transmission is evaluated at 11 uniformly spaced frequencies across this bandwidth, capturing the broadband coupling behavior while maintaining computational feasibility.
At the output, a mode monitor is positioned $18.75~\mu\mathrm{m}$ beyond the inverse-design region within the uniform waveguide and measures the power coupled into the fundamental mode. Positioning the monitor beyond the inverse-design region gives the transmitted field some distance to settle into the uniform waveguide mode before the coupling efficiency is measured.

Within the simulation, the broadband coupling efficiency is defined as the mean fundamental-mode coupling efficiency across the sampled frequencies,
\begin{equation}
   T_{\mathrm{avg}} =
\frac{1}{N}
\sum_{i=1}^{N}
\left|a_i\right|^2 ,
\end{equation}
where $a_i$ is the forward-propagating fundamental-mode amplitude at frequency $i$ and $N=11$ is the number of sampled frequencies. Tidy3D normalizes the frequency-domain monitor data by the source spectrum, removing the Gaussian spectral weighting of the excitation. Thus, $\left|a_i\right|^2$ directly represents the fraction of input power coupled into the fundamental output mode at frequency $i$, and no additional normalization by the input power is required.

\subsubsection{Shape optimization}
\label{subsec:shape_optimization}

For shape optimization, we initialize the optimization process by connecting the fiber and waveguide with a linear taper composed of Ge-doped silica surrounded by silica cladding. The taper boundary is parameterized by 101 points evenly distributed along its $100~\mu\mathrm{m}$ length. At each longitudinal position, a single parameter defines the taper half-width, and the upper and lower boundaries are mirrored about the propagation axis, enforcing a symmetric taper and reducing the number of independent design parameters.

The taper endpoints are fixed to match the $8.2~\mu\mathrm{m}$ fiber-core diameter at the input and the $3.6~\mu\mathrm{m}$ waveguide width at the output, while the 99 interior points are treated as optimization variables. Each interior taper half-width is parameterized by an unconstrained variable that may vary from $-\infty$ to $+\infty$. A hyperbolic-tangent transformation is then used to map this variable to a physically meaningful half-width bounded between the fiber and waveguide half-widths. This bounds each point between the input and output half-widths without requiring explicit clipping during optimization.

Independent movement of the boundary points can produce sharp kinks or rapidly varying features; thus, a curvature penalty is included in the figure of merit to encourage fabricable designs. The penalty discourages local radii of curvature below $0.15~\mu\mathrm{m}$, corresponding to the assumed minimum fabrication scale. The resulting objective can therefore be written schematically as

\begin{equation}
F_{\mathrm{shape}} =
T_{\mathrm{avg}} -
w P_{\mathrm{curv}},
\end{equation}

where $T_{\mathrm{avg}}$ is the mean broadband coupling efficiency, $P_{\mathrm{curv}}$ is the curvature penalty, and $w$ is the penalty weight, which controls the relative strength of the fabrication constraint. Gradient-based optimization then modifies the interior taper parameters to maximize this objective. The shape optimization is performed for 200 iterations using the Adam optimizer \cite{adam} with a learning rate of 0.01 and a penalty weight, $w$, of 0.01. 

\subsubsection{Topology optimization}
\label{subsec:topology_optimization}

Topology optimization relaxes the use of a predefined geometry and instead optimizes the material distribution throughout the fiber-to-waveguide transition. The design region spans $100~\mu\mathrm{m}$ along the propagation direction and $18~\mu\mathrm{m}$ in the transverse direction, corresponding to five times the waveguide width. The region is discretized into $0.3~\mu\mathrm{m}\times0.3~\mu\mathrm{m}$ pixels, each with its own density parameter. These pixels define a two-dimensional density field in the propagation--transverse plane, which is then extended uniformly through the $3.6~\mu\mathrm{m}$ height of the waveguide core. Thus, the material density of each pixel is held constant along the vertical direction. This reflects the intended fabrication process, in which each in-plane pixel is assigned a single material throughout the device height. The $0.3~\mu\mathrm{m}$ pixel size, which is coarser than the assumed $0.15~\mu\mathrm{m}$ fabrication scale, was chosen to reduce the computational cost of the three-dimensional FDTD simulations.

The density of each pixel is represented by a continuous design variable $\rho\in[0,1]$. A value of $\rho=0$ corresponds to silica, while $\rho=1$ corresponds to Ge-doped silica. The initial density distribution is generated randomly to allow broad exploration of the design space.
During optimization, intermediate values of $\rho$ are permitted and represent an interpolation between the permittivities of the two materials. These intermediate values provide a continuous design space for gradient-based optimization but do not correspond to physically realizable materials in the final device. To progressively drive the design toward the two physical materials, the density field is passed through a hyperbolic-tangent projection. The sharpness of this projection is controlled by the parameter $\beta$, which is increased exponentially from 5 to 80 throughout the optimization. At low $\beta$, intermediate density values remain possible, allowing the material distribution to change smoothly as the optimizer explores the design space. As $\beta$ increases, the projection becomes sharper and the density values are increasingly driven toward $\rho=0$ or $\rho=1$, such that the final design approaches a binary distribution of silica and Ge-doped silica.

Before applying this projection, the density field is processed using a conic spatial filter. The conic filter smooths the density field over a prescribed spatial radius, suppressing small-scale variations and encouraging larger, more fabricable features. A filter radius of $0.45~\mu\mathrm{m}$ is used, which smooths the density field over a length scale larger than both the $0.3~\mu\mathrm{m}$ pixel size and the assumed $0.15~\mu\mathrm{m}$ minimum fabrication scale.

An erosion--dilation penalty is additionally applied to discourage narrow or fragile features that may be sensitive to fabrication errors. The contribution of this penalty is gradually increased during optimization as the projection becomes more binary. The resulting objective is

\begin{equation}
F_{\mathrm{topology}} =
T_{\mathrm{avg}} -
P_{\mathrm{ED}},
\end{equation}

where $T_{\mathrm{avg}}$ is the mean broadband coupling efficiency and $P_{\mathrm{ED}}$ is the erosion--dilation penalty.

The topology optimization is performed for 100 iterations using the Adam optimizer \cite{adam} with a learning rate of 0.1. We choose a lower number of iterations than shape optimization due to the higher computational cost of each topology optimization simulation. At each iteration, gradients of the objective function with respect to the pixel densities are calculated and used to update the material distribution. The design variables are constrained to remain within $[0,1]$ throughout the optimization.

The code for this work is available through Github \href{https://github.com/Astrophotonics-Technology-Lab/seenivasan_SPIE_2026}{here}. 

\section{Results}

\subsection{Shape Optimization}

\begin{figure}[htbp]
    \centering

    \begin{subfigure}[t]{0.32\textwidth}
        \centering
        \includegraphics[width=\linewidth]{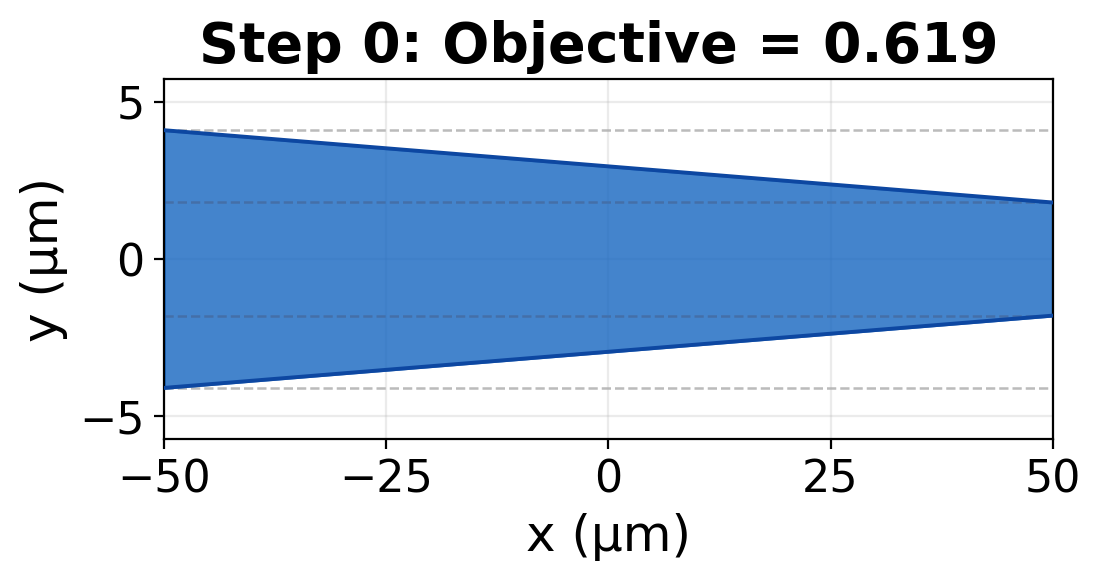}
        \caption{Initial}
        \label{fig:shape_initial}
    \end{subfigure}
    \hfill
    \begin{subfigure}[t]{0.32\textwidth}
        \centering
        \includegraphics[width=\linewidth]{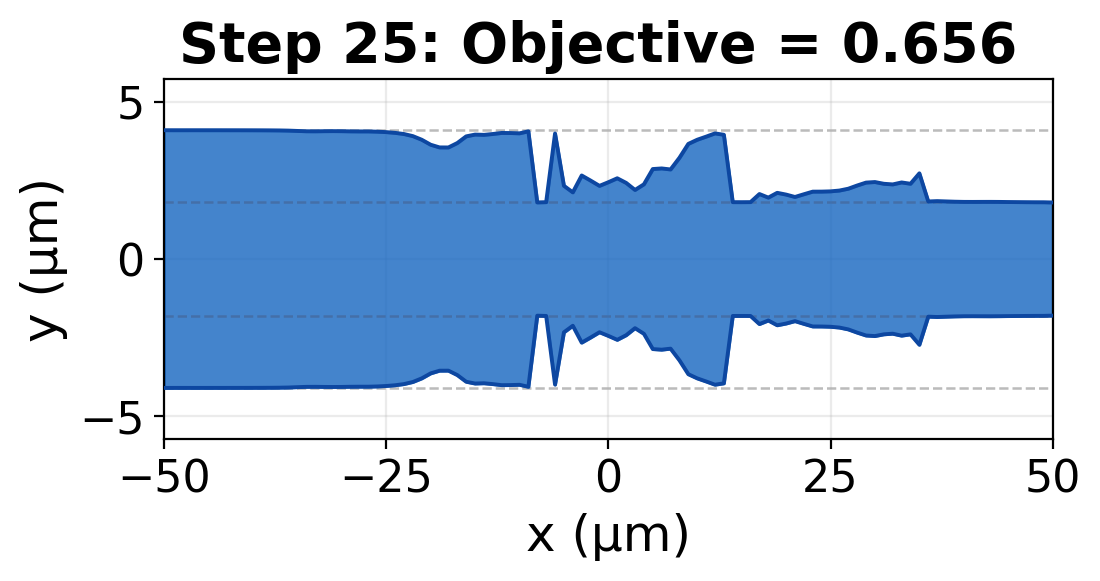}
        \caption{Intermediate}
        \label{fig:shape_middle}
    \end{subfigure}
    \hfill
    \begin{subfigure}[t]{0.32\textwidth}
        \centering
        \includegraphics[width=\linewidth]{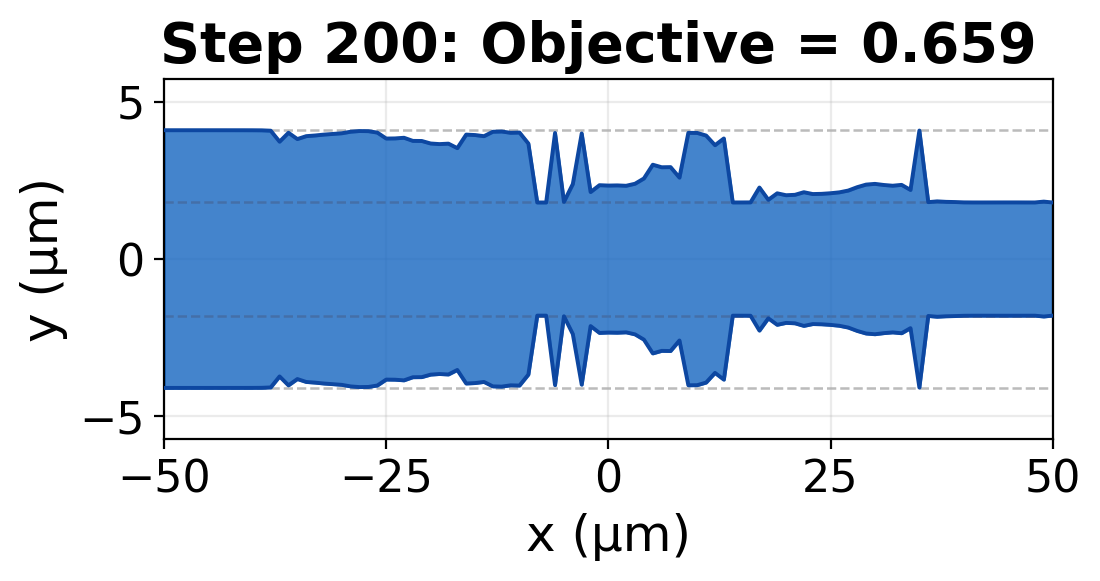}
        \caption{Final}
        \label{fig:shape_final}
    \end{subfigure}

    \caption{
    Evolution of the shape-optimized fiber-to-waveguide transition.
    The taper shape is shown from left to right at
    (a) initialization at step 0,
    (b) an intermediate optimization step, 25, and
    (c) the final optimization step, 200.
    }
    \label{fig:shape_evolution}
\end{figure}

\begin{figure}[htbp]
    \centering

    \begin{minipage}[t]{0.35\textwidth}
        \centering
        \includegraphics[width=\linewidth]{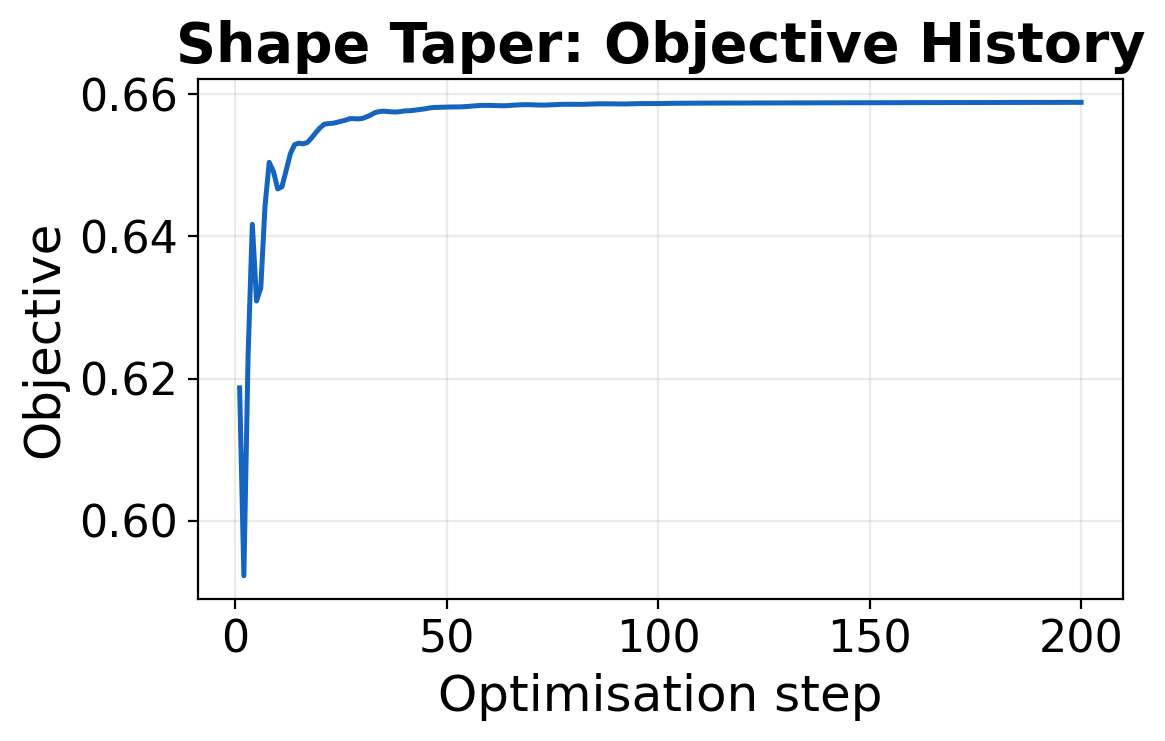}
        \caption{Evolution of the shape optimization objective.}
        \label{fig:shape_objective_history}
    \end{minipage}
    \hspace{0.03\textwidth}
    \begin{minipage}[t]{0.35\textwidth}
        \centering
        \includegraphics[width=\linewidth]{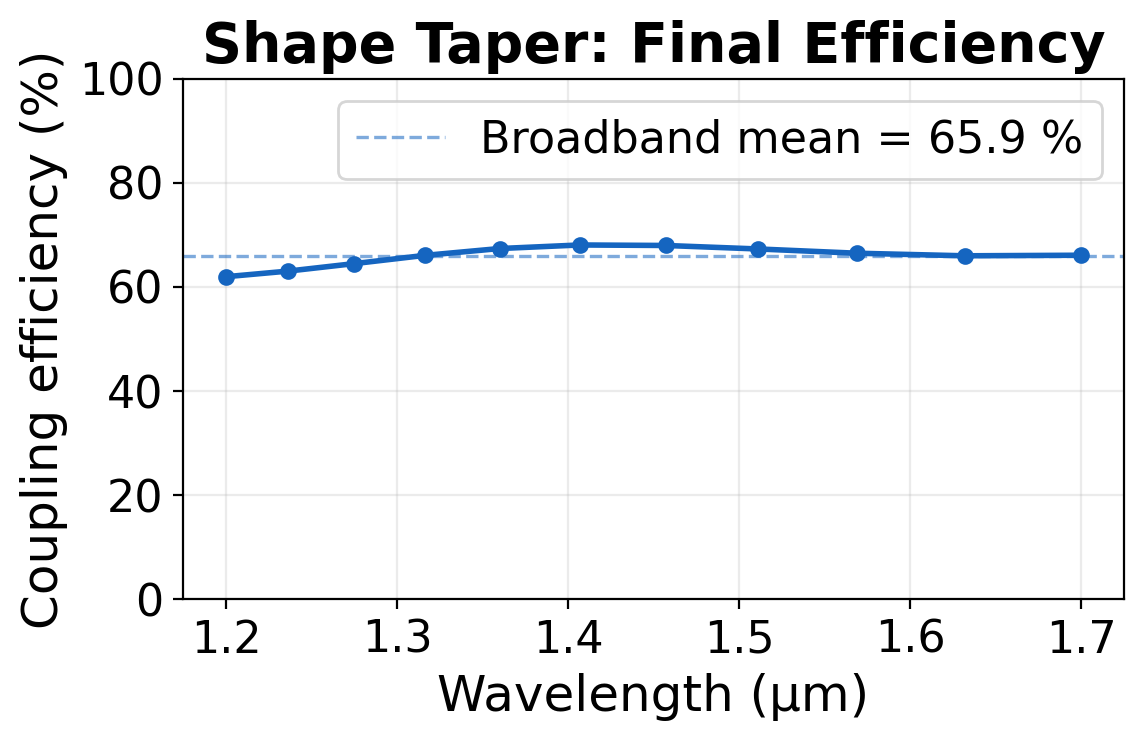}
        \caption{The final broadband efficency of the shape-optimized taper.}
        \label{fig:shape_efficiency}
    \end{minipage}
    
\end{figure}

Fig. \ref{fig:shape_evolution} shows the evolution of the taper geometry during shape optimization. The objective increases from 0.619 at Step 0 to 0.656 at Step 25, and finally reaching 0.659 at Step 200. The optimized shape of the taper boundary has a stepped profile with three approximately rectangular sections, with progressively smaller  heights. There are several spike-like features in the final taper geometry. Fig. \ref{fig:shape_objective_history} shows the shape optimization objective over the 200 optimization steps. The objective increases rapidly from 0.619 and approaches its final value of 0.659 within $\sim$50 steps, and remains relatively constant until the end of the optimization. 
Fig. \ref{fig:shape_efficiency} shows the final broadband efficiency of the taper across the 1200-1700 nm wavelength range. The coupling efficiency is fairly uniform across all wavelengths, with a mean value of 65.9\%. 

\subsection{Topology Optimization}

\begin{figure}[htbp]
    \centering

    \begin{subfigure}[t]{0.32\textwidth}
        \centering
        \includegraphics[width=\linewidth]{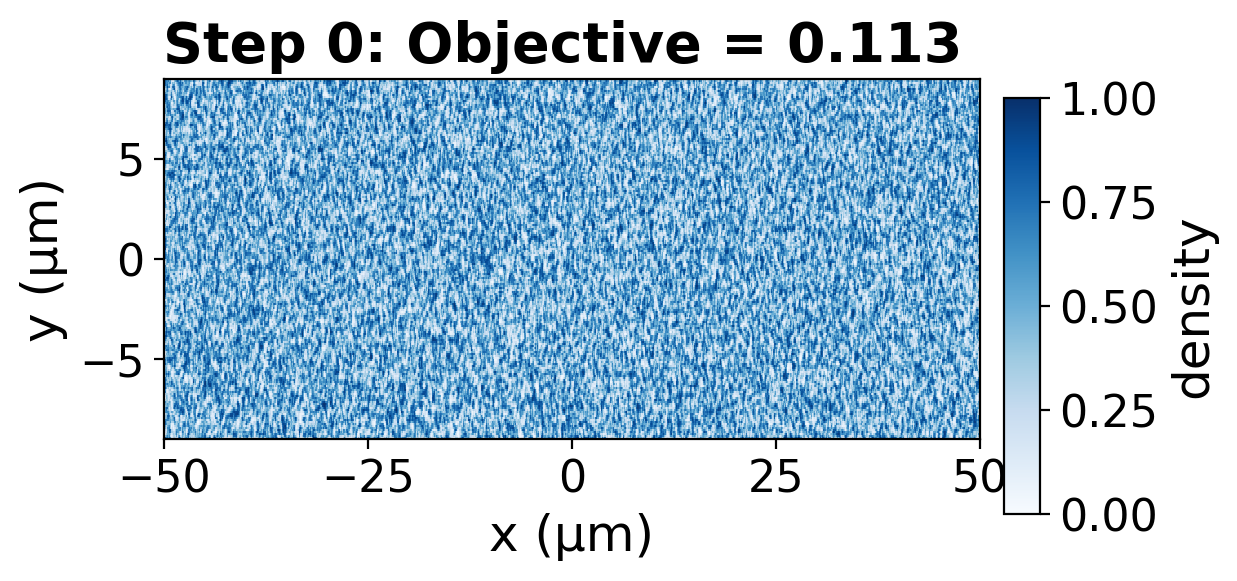}
        \caption{Initial}
        \label{fig:topology_initial}
    \end{subfigure}
    \hfill
    \begin{subfigure}[t]{0.32\textwidth}
        \centering
        \includegraphics[width=\linewidth]{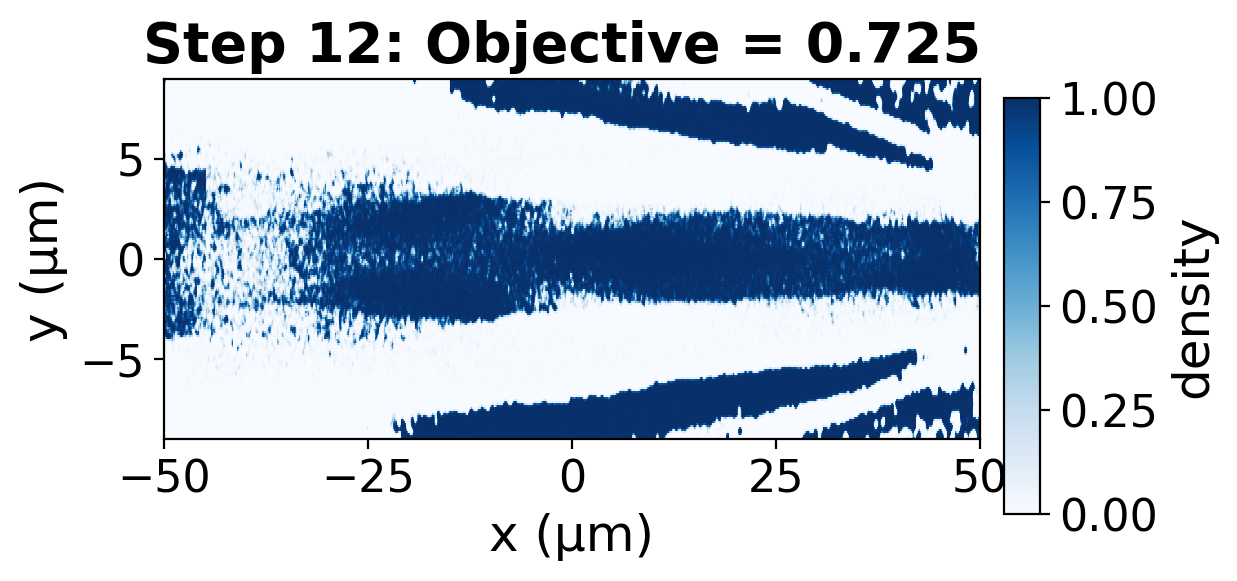}
        \caption{Intermediate}
        \label{fig:topology_middle}
    \end{subfigure}
    \hfill
    \begin{subfigure}[t]{0.32\textwidth}
        \centering
        \includegraphics[width=\linewidth]{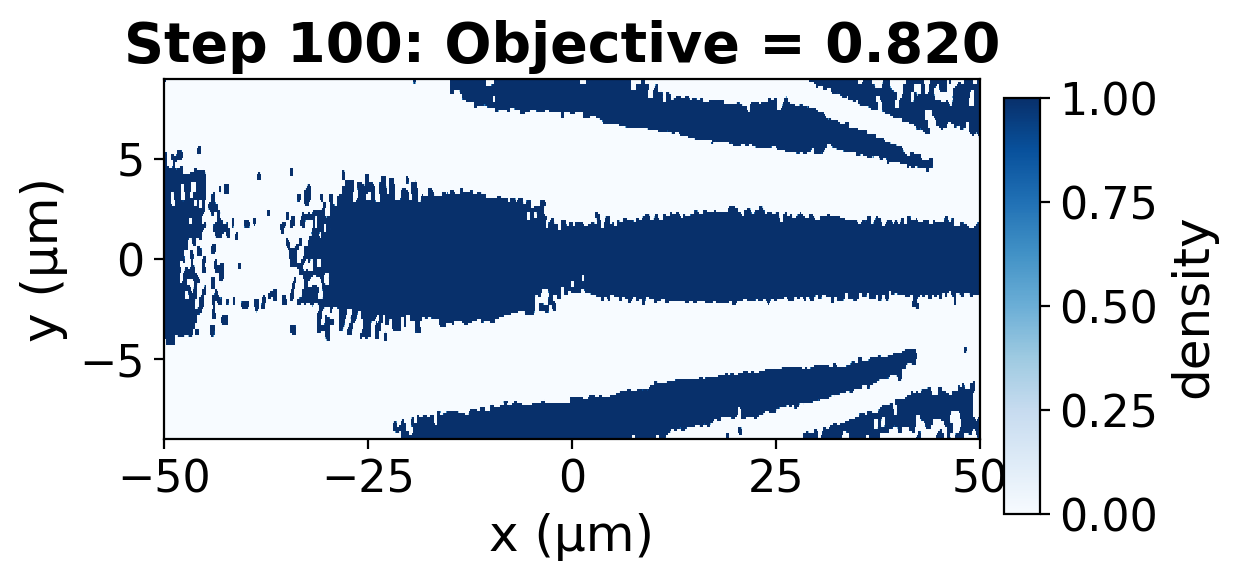}
        \caption{Final}
        \label{fig:topology_final}
    \end{subfigure}

    \caption{
    Evolution of the topology-optimized fiber-to-waveguide transition.
    The material distribution and binarization process is shown from left to right at
    (a) initialization, step 0,
    (b) an intermediate optimization step, 12, and
    (c) the final optimization step, 100.
    }
    \label{fig:topology_evolution}
\end{figure}

\begin{figure}[htbp]
    \centering

    \begin{minipage}[t]{0.35\textwidth}
        \centering
        \includegraphics[width=\linewidth]{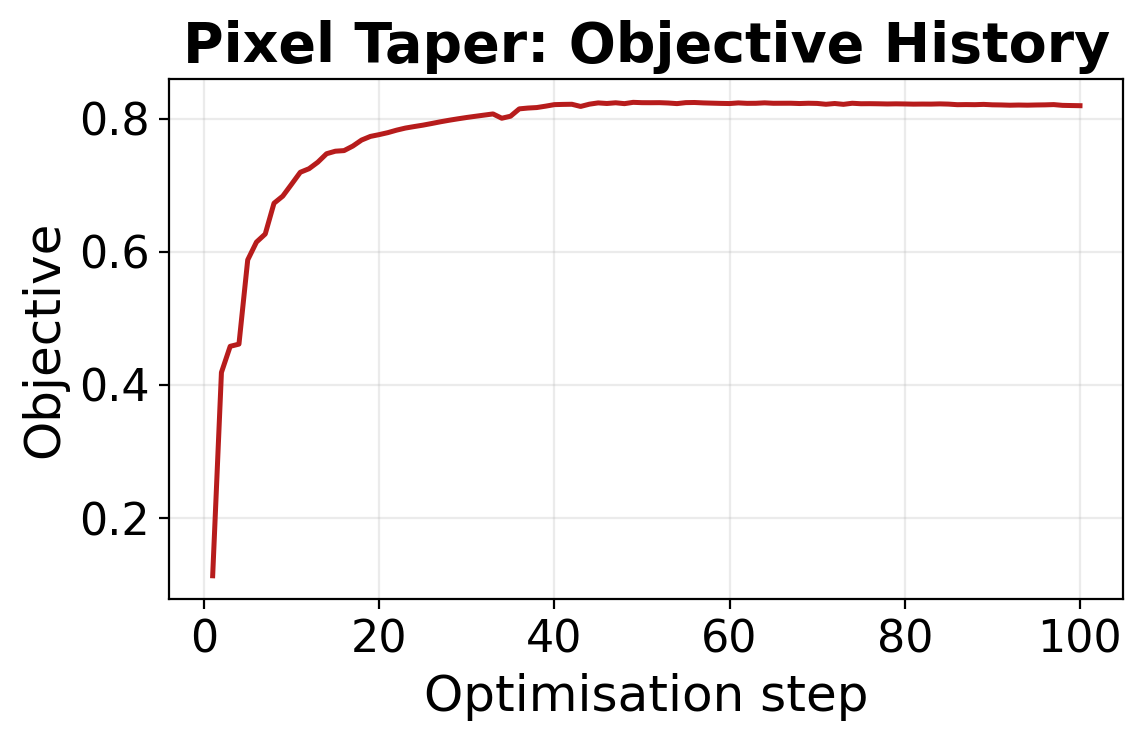}
        \caption{Evolution of the topology optimization objective.}
        \label{fig:topology_objective_history}
    \end{minipage}
    \hspace{0.03\textwidth}
    \begin{minipage}[t]{0.35\textwidth}
        \centering
        \includegraphics[width=\linewidth]{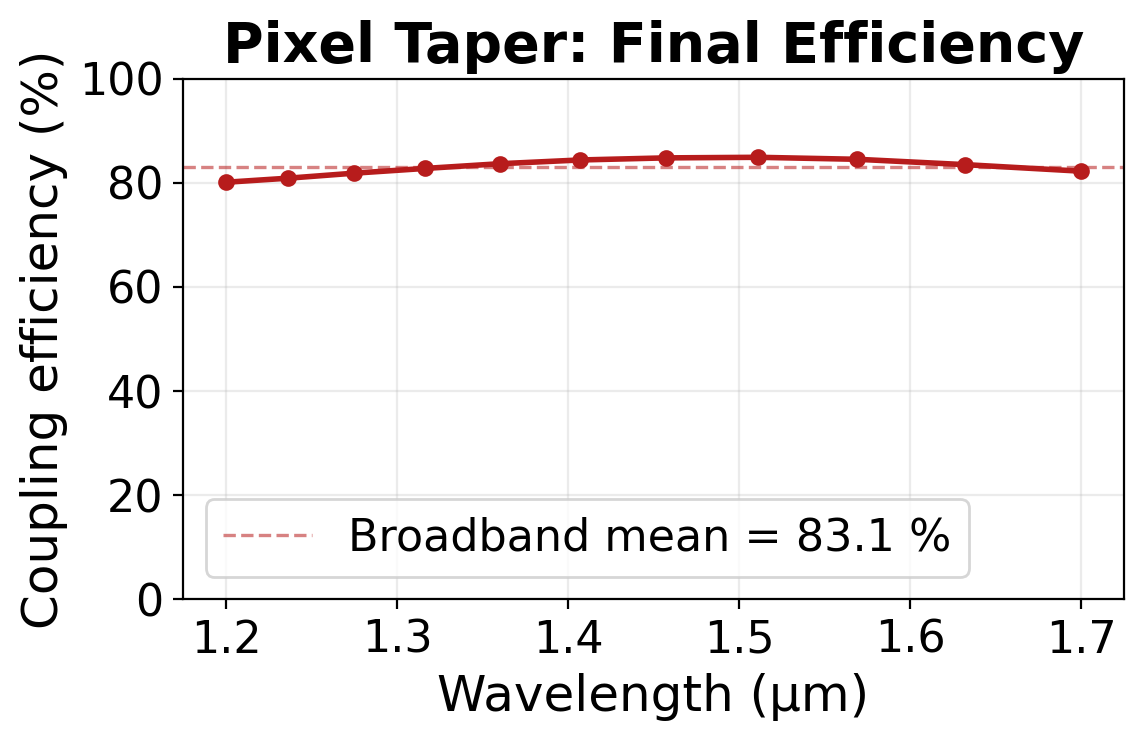}
        \caption{The final broadband efficiency of the topology-optimized interface.}
        \label{fig:topology_efficiency}
    \end{minipage}

\end{figure}

Fig. \ref{fig:topology_evolution} shows the evolution of the material distribution in the design region during topology optimization. The objective increases rapidly from 0.113 at Step 0 to 0.725 by Step 12 and gradually reaches 0.820 at Step 100. The optimized material distribution develops a taper-like region of Ge-doped silica that is separated from the input SMF-28 fiber by a predominantly silica region. Some sparsely distributed Ge-doped silica is present near the SMF-28 fiber input. Additional bands of Ge-doped silica form above and below the central taper-like structure. Figure~\ref{fig:topology_objective_history} shows the topology-optimization objective over the 100 optimization steps. The objective increases rapidly from 0.113 to approximately 0.8 within the first 40 steps and subsequently stabilizes around this value until the end of optimization. 
Figure~\ref{fig:topology_efficiency} shows the broadband coupling efficiency of the final topology-optimized design over the 1200--1700~nm wavelength range. The coupling efficiency is relatively uniform across the full bandwidth, with a mean coupling efficiency of 83.1\%. 

\section{Discussion \& Conclusion}

This preliminary study demonstrates proof-of-concept of inverse design for improving astrophotonic spectrograph throughput. Substantial further work is required to reach the efficiencies needed for practical deployment. 

The results highlight a tradeoff between the two inverse-design approaches. Shape optimization improves the broadband coupling efficiency by only $\sim4\%$ relative to the baseline linear taper. Its restricted parameterization limits the range of geometries that can be explored, since the optimization is confined to modifying the taper boundary points. The narrow spike-like features that remain in the shape-optimized taper also suggest that increasing the fabrication-penalty weight could lead to more robust, manufacturable designs.
Topology optimization, on the other hand, achieves a broadband coupling efficiency $\sim17.2\%$ higher than the shape-optimized taper. This improvement may be due to its substantially greater design freedom: rather than modifying only the taper boundary, topology optimization independently adjusts the material distribution across many pixels throughout the design region. Within the approximations and computational limits of this preliminary study, these results suggest that topology optimization may offer greater potential for improving coupling efficiency in astrophotonic spectrographs.

A major limitation of the present study is the computational cost of three-dimensional electromagnetic simulations. This restricts the resolution and physical extent of the design region, whether or not the material dispersion is included in the simulations, the number of frequencies that can be simulated, as well as the number of optimization configurations that can be explored. In particular, we did not perform a search over hyperparameters such as learning rate, iteration count, fabrication-penalty strength, filter radius, or projection schedule, nor did we systematically investigate different design-region lengths or initial conditions. Future work will therefore focus on more thoroughly investigating the topology-optimization design space, including hyperparameter optimization, larger and higher-resolution design regions, alternative initializations, material dispersion, and finer frequency sampling, to determine whether the present coupling efficiency can be further improved.

The inverse-design framework will subsequently be extended to the other interfaces throughout the spectrograph, followed by fabrication and experimental characterization of the optimized device. Ultimately, achieving an overall spectrograph throughput of 95\% or higher, an ambitious goal for high-precision astronomical applications, will require individual interfaces to approach approximately 99\% efficiency. Inverse design potentially offers a promising pathway toward this goal by enabling the discovery of efficient, non-intuitive photonic structures. 

\acknowledgments       
 
The authors thank Samantha Morisson, Sydney Skorpen and Srinath Saikrishnan for their advice and support in conducting this research and preparing this manuscript. The authors also thank Filipe Cuoto from Flexcompute for providing technical support for the Tidy3D simulation and inverse-design framework. 
This work was supported by the National Aeronautics and Space Administration (NASA) under grant number 80NSSC24K1560, awarded through the Astrophysics Research and Analysis (APRA) program. In addition, we acknowledge support from the Kavli Foundation through the Kavli Exploration Award.

\bibliography{report} 
\bibliographystyle{spiebib} 

\end{document}